\documentstyle[prl,aps,epsf]{revtex}
\begin{document}
\newcommand{\be}{\begin{equation}}
\newcommand{\ee}{\end{equation}}
\newcommand{\bc}{\begin{center}}
\newcommand{\ec}{\end{center}}
\newcommand{\bi}{\begin{itemize}}
\newcommand{\ei}{\end{itemize}}
\newcommand{\bo}{\={o} }
\newcommand{\th}{$\Theta$}

\title{Discrete molecular dynamics studies of\\ the folding of a
protein-like model}
\author{Nikolay V. Dokholyan, Sergey V. Buldyrev, H. Eugene Stanley}
\address{Center for Polymer Studies, Physics Department, Boston University,
Boston, MA 02215, USA}
\author{Eugene I. Shakhnovich\footnote{To whom correspondence should be
addressed; email: eugene@diamond.harvard.edu}}
\address{Department of Chemistry, Harvard University, 12 Oxford Street,
Cambridge, MA 02138, USA}

%\date{\today ~~~ draft}
\bigskip
\date{Folding \& Design, {\em in press}}
\maketitle

\bigskip

\begin{abstract}
\noindent {\bf Background:} Many attempts have been made to resolve in time the
folding of model proteins in computer simulations. Different computational
approaches have emerged. Some of these approaches suffer from the
insensitivity to the geometrical properties of the proteins (lattice
models), while others are computationally heavy (traditional MD).\\
{\bf Results:} We use a recently-proposed approach of Zhou and Karplus
to study the folding of the protein model based on the discrete time
molecular dynamics algorithm. We show that this algorithm resolves with
respect to time the folding $\rightleftharpoons$ unfolding
transition. In addition, we demonstrate the ability to study the coreof
the model protein.\\
{\bf Conclusion:} The algorithm along with the model of inter-residue
interactions can serve as a tool to study the thermodynamics and kinetics
of protein models.
\end{abstract}

\section{Introduction}

The vast dimensionality of the protein conformational space
\cite{Levinthal68} makes the folding time too long to be reachable by
direct computational approaches \cite{Go83,Karplus94,Shakhnovich97}.
Simplified models
\cite{Taketomi75,Go81,Abe81,Dill85,Bryngelson89,Shakhnovich94,Abkevich94,Gutin95,Shakhnovich96,Dill90}
became popular due to their ability to reach reasonable time scales and
to reproduce the basic thermodynamic and kinetic properties of real
proteins \cite{Karplus94,Creighton92,Privalov89}: {\it (i)} unique
native state, i.e. there should exist a single conformation with the
lowest potential energy; {\it (ii)} cooperative folding transition
(resembling first order transition); {\it (iii)} thermodynamical
stability of the native state; {\it (iv)} kinetic accessibility,
i.e. the native state should be reachable in a biologically reasonable
time \cite{Gutin95,Klimov96}.

Monte Carlo (MC) simulations on the lattices (see, e.g.,
\cite{Shakhnovich97,Taketomi75,Go81,Abe81} and references therein)
appear to be useful for studying theoretical aspects of protein folding. 
The Monte Carlo algorithm is based on a set of rules for the
transition from one conformation to another. These transitions are
weighted by some transition matrix, which reflects the phenomena under
study. The simplicity of the algorithm and a significantly small
conformational space of the protein models (due to the lattice
constraints) make MC on-lattice simulations a powerful tool for studying
the equilibrium dynamics of the protein models. However, lattice models
impose strong constraints on the angles between the covalent bonds,
thereby greatly restricting the conformational space of the protein-like
model. The additional drawback of this restriction lies in the poor
capability of these models to discern the geometrical properties of the
proteins. The time in MC algorithms is estimated as the average number
of moves (over an ensemble of the folding $\rightleftharpoons$ unfolding
transitions) made by a model protein. It was pointed out 
\cite{Baumgartner87} that MC simulations are equivalent to the solution 
of the master equation for the dynamics, so there is a relation
between physical time and computer time, which is counted as the number
of MC steps. However, a number of delicate issues --- such as the
dependence of the dynamics on the set of allowed MC moves --- remain
outstanding, so an independent test of the dynamics using the MD
approach is needed.

To address the questions sensitive to geometrical details, it is useful
to study off-lattice models of protein folding. Thus far, several
off-lattice simulations have been performed
\cite{Irback95,Berriz97,Guo97}, which demonstrate the ability of the
simplified models to study protein folding.

Here, we study the 3-dimensional molecular dynamics of a simplified
model of proteins \cite{Go81,Abe81}. The potential of interaction
between pairs of residues is modeled by a ``square-well'', which allows
us to increase the speed of the simulations \cite{Zhou97,Zhou97a}. We
estimate folding time based on the collision event list, which besides
increasing the speed of the simulation, allows for the tracking of
``realistic'' (not discretized) time. We show that such an algorithm can
be a useful compromise between computationally heavy traditional MD and
fast, but restrictive MC. We demonstrate that model protein reproduces
the principal features of folding phenomena {\it (i) -- (iv)} described
above.

We also address the question of whether we can study the equilibrium
properties of the core. {\em The core is a small subset of the residues,
which maintains the backbone of the structure at temperatures close to
the folding transition temperature (here the \th-temperature
$T_{\theta}$)}. We emphasize the difference between the core and the
nucleus of a protein: while the core is a persistent part of the
structure at equilibrium, the nucleus is a fragment of this structure,
which is assembled in the transition state (TS) --- the folding
$\rightleftharpoons$ unfolding barrier (see Fig.~1 in
\cite{Shakhnovich97}). Based on simple arguments, we estimate
$T_{\theta}$\cite{Doi96} for our model, and compare it with the value
found in the simulations.

\section{The model}
\begin{figure}[htb]
\centerline{
\epsfxsize=8.0cm
\epsfbox{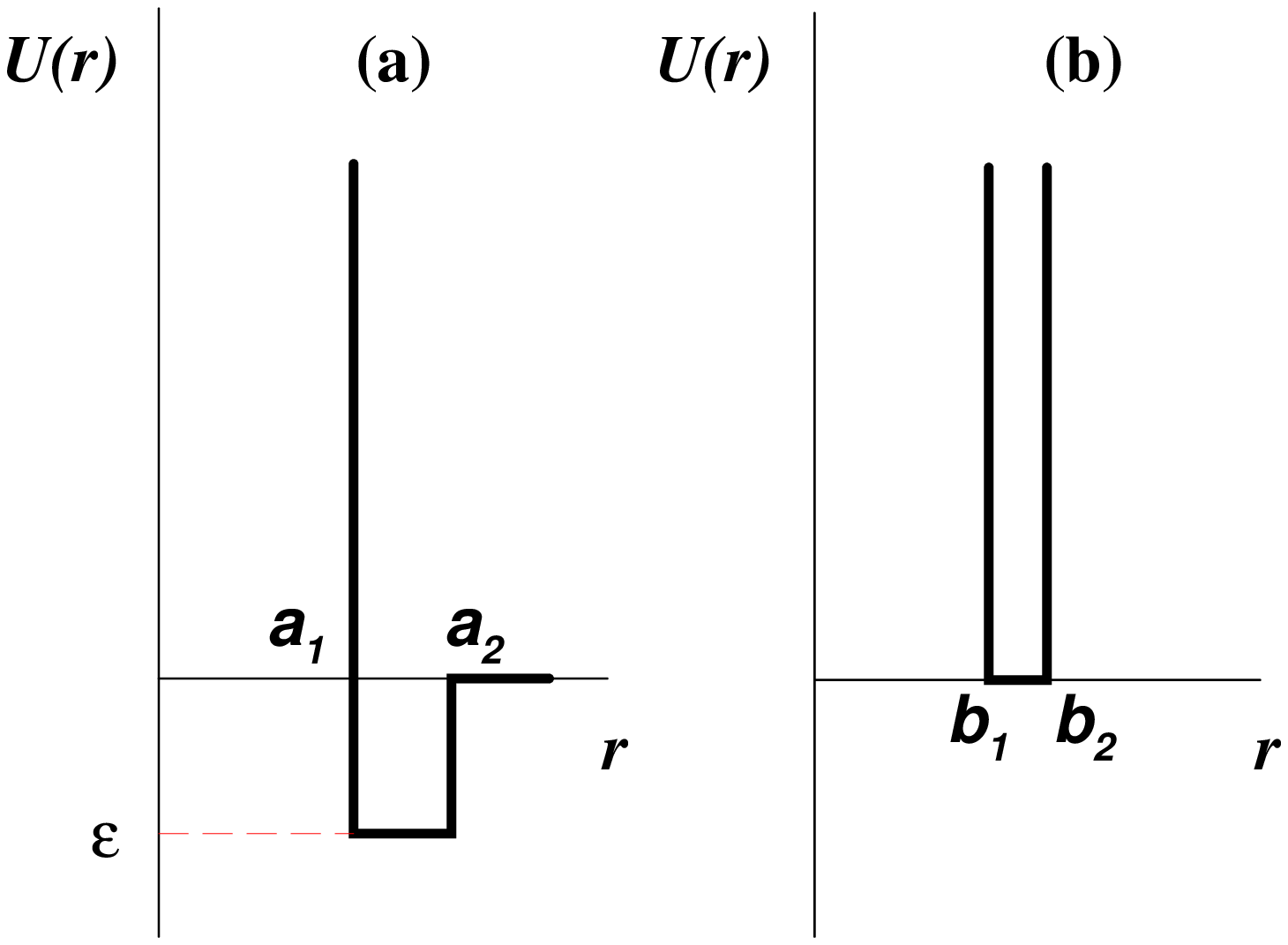}
\vspace*{1.0cm}
}
\caption{The potential of interaction between (a) specific residues; (b)
neighboring residues (covalent bond). $a_0$ is the diameter of the
hard sphere and $a_1$ is the diameter of the attractive sphere. $[b_0,
b_1]$ is the interval where residues that are neighbors on the chain can
move freely.}
\label{fig:1}
\end{figure}

\begin{figure}[htb]
\centerline{
\epsfxsize=8.0cm
\epsfbox{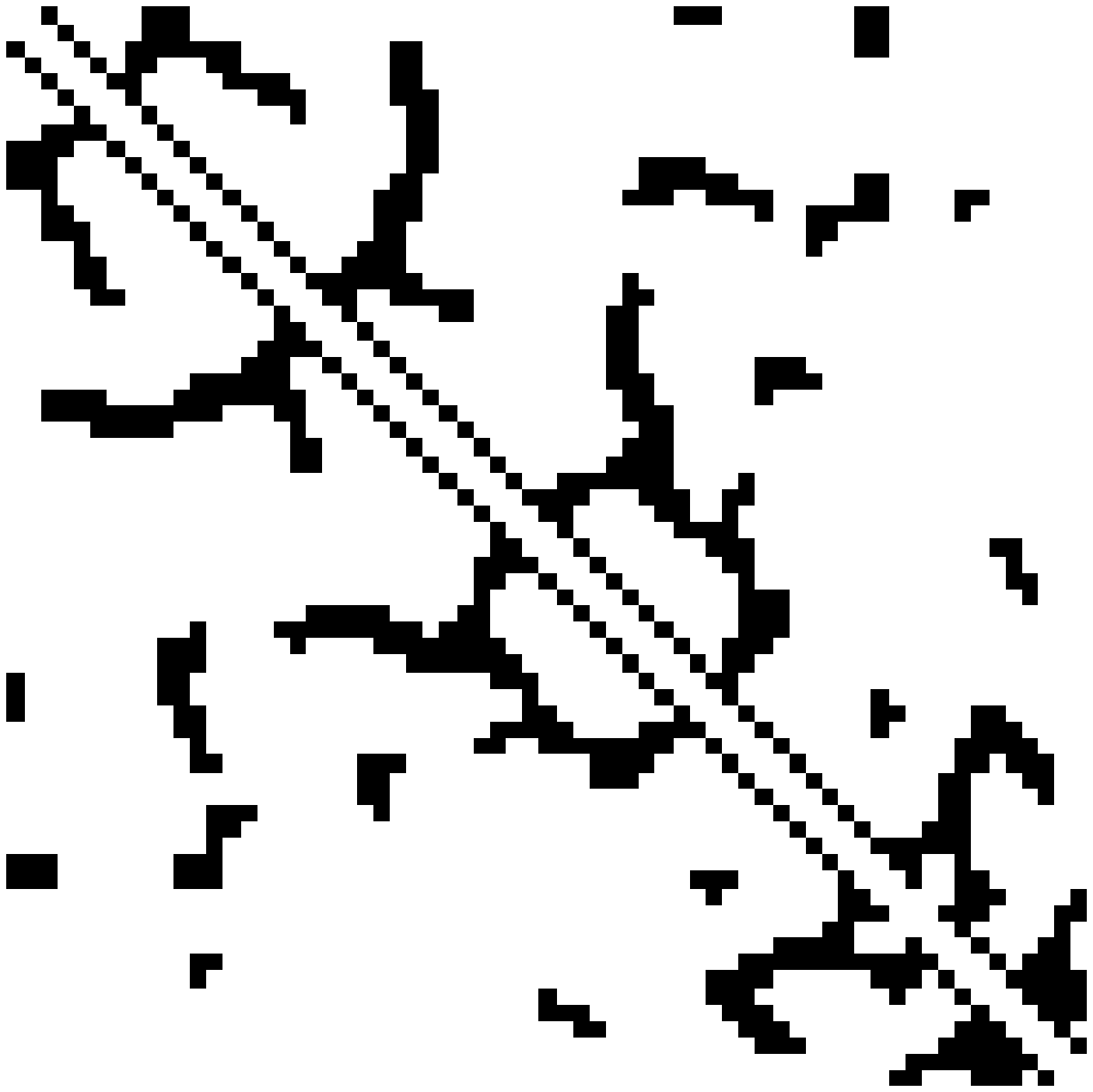} 
\vspace*{1.0cm}
%}
%\centerline{
\epsfxsize=8.0cm
\epsfbox{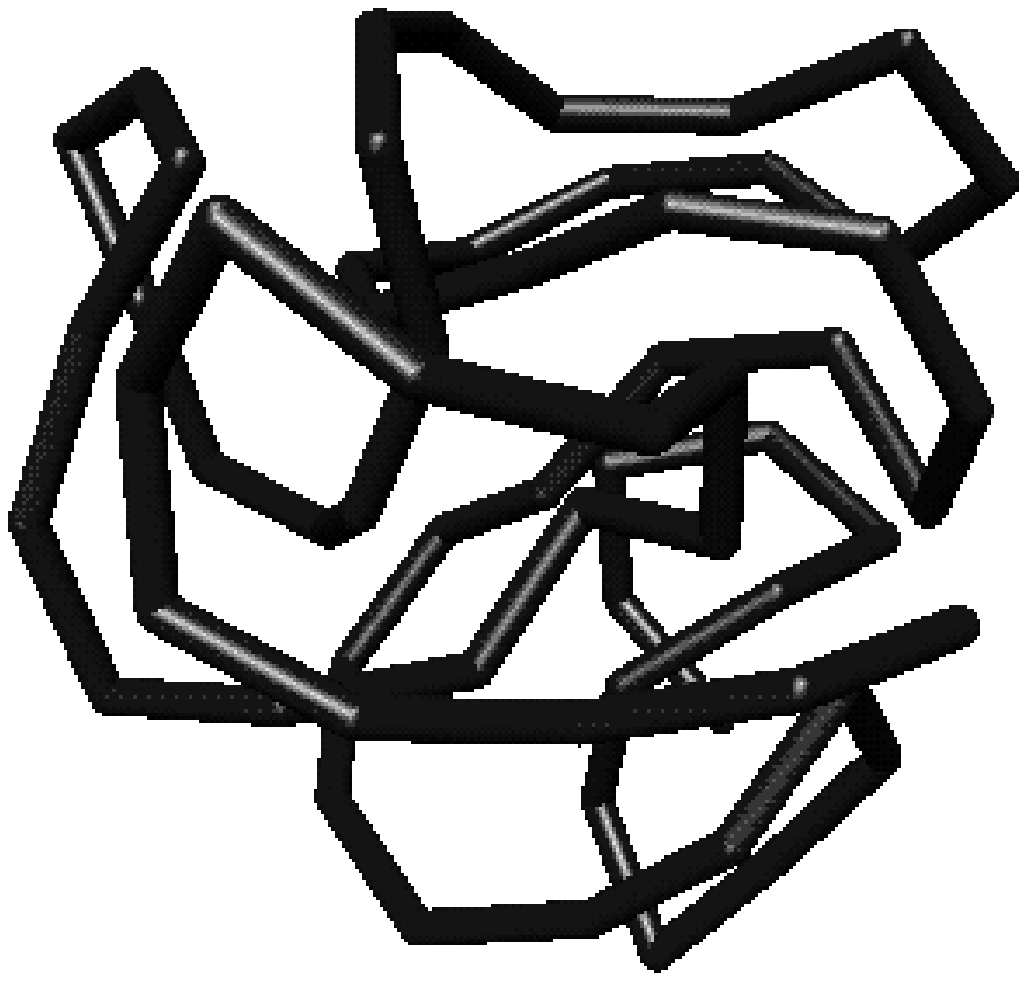}
\vspace*{1.0cm}
}
\caption{(a) 65 x 65 contact matrix of the model protein in the native state.
Black boxes  indicate the matrix elements of those residue pairs which
have a contact (their relative distance is between $a_0$ and $a_1$). (b)
The snapshot of the protein of 65 residues in the native state obtained at
temperature $T=0.1$.}
\label{fig:2}
\end{figure}

We study a ``beads on a string'' model of a protein. We model the
residues as hard spheres of unit mass. The potential of interaction
between residues is ``square-well''. We follow the G\bo model
\cite{Taketomi75,Go81,Abe81}, where the attractive potential between
residues is assigned to the pairs that are in contact ($\Delta_{ij}$,
defined below) in the native state and repulsive potential is assigned
to the pairs that are not in contact in the native state. Thus, the
potential energy
\be
{\cal E} = \frac{1}{2}\sum_{i,j=1}^{N}{U_{i,j}}\, \, 
\label{eq:U}
\ee
where $i$ and $j$ denote residues $i$ and $j$. $U_{i,j}$ is the
matrix of pair interactions
\be
       U_{i,j} = \left\{ \begin{array}{ll}
                   +\infty , & |r_i-r_j|\le a_0\\
                   -\mbox{sign}(\Delta_{ij})\epsilon, &a_0<|r_i-r_j|\le a_1\\  
                   0, &  |r_i-r_j| > a_1\, .
                  \end{array} \label{eq:Uij}
\right.
\ee
Here $a_0/2$ is a radius of the hard sphere, and $a_1/2$ is the radius of
the attractive sphere (Fig.~\ref{fig:1}a) and $\epsilon$ sets the energy
scale. $||\Delta||$ is a matrix of contacts with elements
\be
\Delta_{ij} \equiv \left\{ \begin{array}{ll}
                   1, &  |r_i^{NS}-r_j^{NS}|\le a_1\\  
                  -1, &  |r_i^{NS}-r_j^{NS}| > a_1\, ,
                  \end{array} \label{eq:Delta}
\right.
\ee
where $r_i^{NS}$ is the position of the $i^{th}$ residue when the
protein is in the native conformation. Note, that we penalize the
non-native contacts by imposing $\Delta_{ij}<0$. The parameters are
chosen as follows: $\epsilon=1$, $a_0=9.8$ and $a_1=19.5$. The covalent
bonds are also modeled by a square-well potential (Bellemans' bonds):
\be
      V_{i,i+1} = \left\{ \begin{array}{ll}
                   0, &  b_0 < |r_i-r_{i+1}| < b_1\\
                   +\infty, &  |r_i-r_{i+1}|\le b_0, \mbox{ or } 
                               |r_i-r_{i+1}|\ge b_1\, .  
                  \end{array} \label{eq:Vnn}
\right.
\ee
The values of $b_0=9.9$ and $b_1=10.1$ are chosen so that average
covalent bond length is equal to 10 (See Fig.~\ref{fig:1}b). The
original configuration of the protein ($N=65$ residues) was designed by
collapse of a homopolymer at low temperature
\cite{Berriz97,Shakhnovich93,Shakhnovich96a}. It contains $n^* = 328$
native contacts, so ${\cal E}_{NS} =-328$. The 65 by 65 matrix of
contacts of the globule in the native state is shown in
Fig.~\ref{fig:2}a. Note that the large number of native contacts
($328/65\approx 5$ contacts per residue) is due to the choice of the
parameter: $a_1\approx 2a_0$ --- so that residues are able to establish
contacts with the residues in the second neighboring shell. The radius
of gyration of the globule in the native state is $R_G\approx 22.7$. The
snapshot of the globule in the native state is shown in
Fig.~\ref{fig:2}b.

The program employs the discrete MD algorithm, which is based on the
collision list, and is similar to one recently used by Zhou et al
\cite{Zhou97} to study equilibrium thermodynamics of homopolymers and by
Zhou and Karplus \cite{Zhou97a} to study equilibrium thermodynamics of
folding of model of {\em Staphylococcus aureus} protein A. The
detailed description of the algorithm can be found in
\cite{Alder59,Grosberg97,Allen87,Rapaport97}. To control the temperature
of the protein we introduce 935 particles, which do not interact with
protein or with each other in any way but via regular collisions,
serving as a heat bath. Thus, by changing the kinetic energy of those
``ghost'' particles we are able to control the temperature of the
environment. The ``ghost'' particles are hard spheres of the same radii
as the chain residues and have unit mass. Temperature is measured in
units of $\epsilon/k_B$. The time unit (tu) is estimated from the
shortest time between two consequent collisions in the system between
any two particles.

\section{Results}

\begin{figure}[htb]
\centerline{
\epsfxsize=8.0cm
\epsfbox{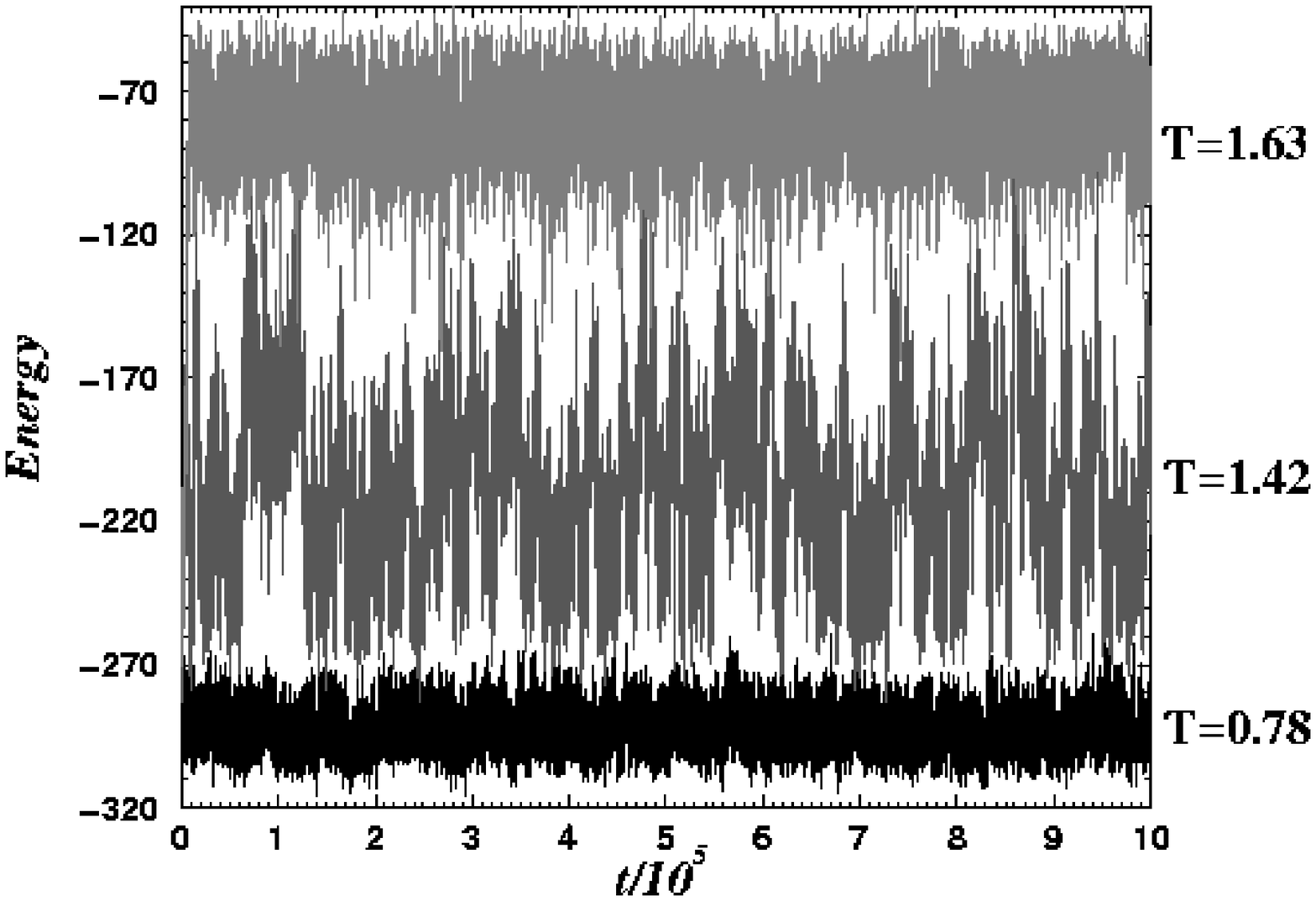}
\vspace*{1.0cm}
%}
%\centerline{
\epsfxsize=8.0cm
\epsfbox{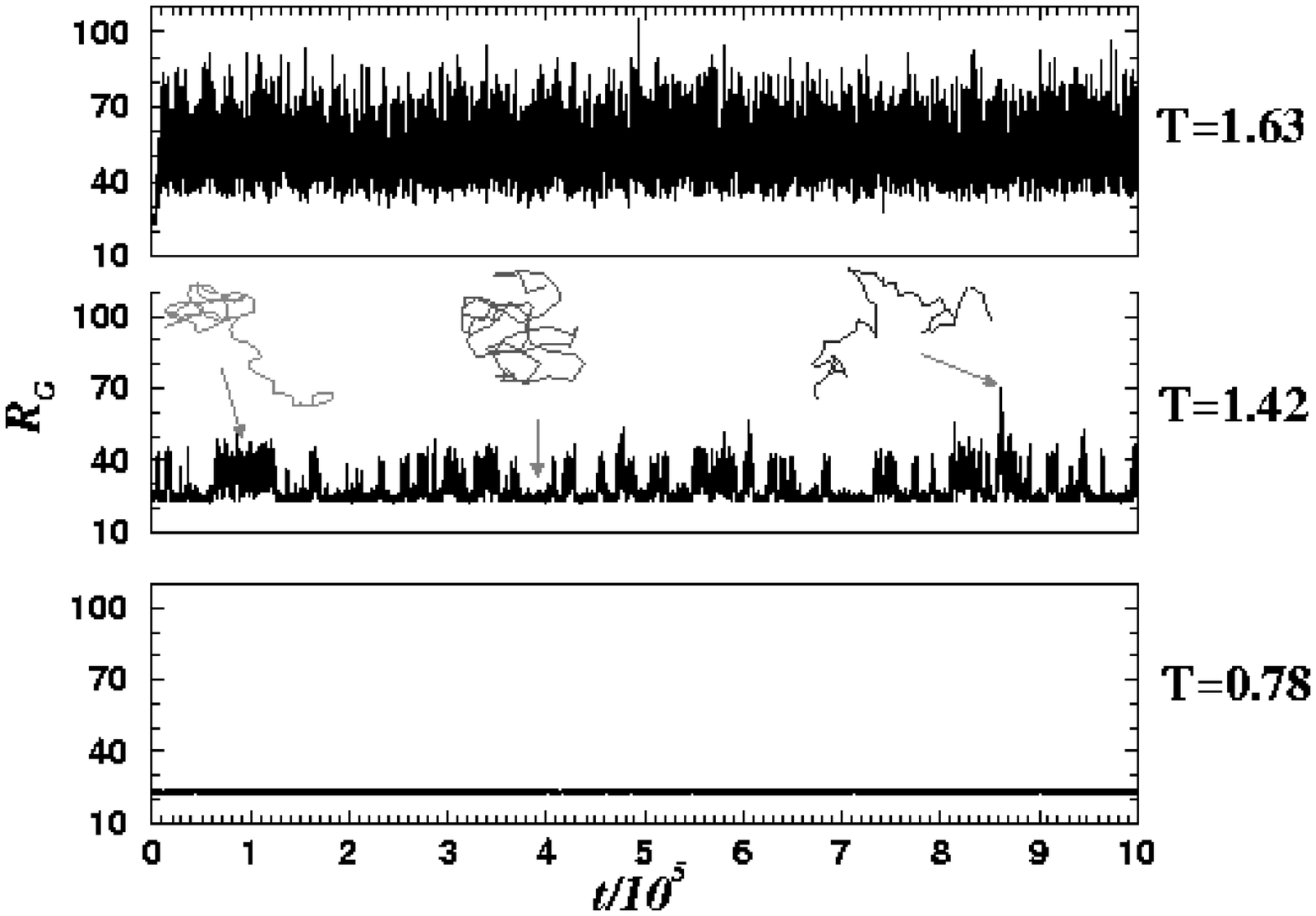}
\vspace*{1.0cm}
}
\caption{The dependence on time of (a) energy ${\cal E}$ and (b) radius
of gyration $R_G$. The globule is maintained at three different
temperatures $T=0.78 <T_f$, $T=1.42$, and $T=1.63 >T_f$ for $10^6$
tu. For $T=0.78$, the fluctuations of both energy ${\cal E}$ 
and $R_G$ are small, i.e. the globule is found in one folded
configuration. At high temperatures ($T=1.63$) the fluctuations of
${\cal E}$ and $R_G$ are large; the globule is mostly found in the
unfolded state. At the temperature $T=1.42$, which is close to $T_f$,
the globule is mostly present in two states. The lower energy
configuration corresponds to the folded state: the globule is compact
(see (b)). The other configuration has large fluctuations: the globule
is in the PFS. There is an additional state -- the unfolded state (see
(b)). At $T=1.42$ the protein model is rarely present in the unfolded
state. Thus, the behavior of the globule at the temperatures close to
$T_f$ indicates the presence of three distinct states: folded, unfolded
and PFS.}
\label{fig:3}
\end{figure}

In order to study the thermodynamics, we perform MD simulations of the
chain at various temperatures. We start with the globule in the native
state at temperature $T=0.1$ and then raise the temperature of the heat
bath to the desired one. Then we allow the system to equilibrate. At the
final temperature, we let the protein relax for $10^6$ time units. The
typical behavior of the energy ${\cal E}$ and the radius of gyration
$R_G$ as functions of time is shown in Fig.~\ref{fig:3} for three
different temperatures.

In the present model the non-native contacts (NNC) are penalized
(i.e., the pairwise interaction between NNC is repulsive\footnote{This
corresponds to $g=2$ of the Ref.~\cite{Zhou97a}.}), so their number
increases as the temperature increases. At high temperatures (above
$T_{\theta}$), however, the number of NNC varies only due to the random
motion of the ideal chain and, thus, on average their number should be
constant at different temperatures. The maximal number of NNC occurs at
$T_{\theta}$ and does not exceed 35, which is roughly 10\% of the total
number of native contacts (NC).

\begin{figure}[htb]
\centerline{
\epsfxsize=8.0cm
\epsfbox{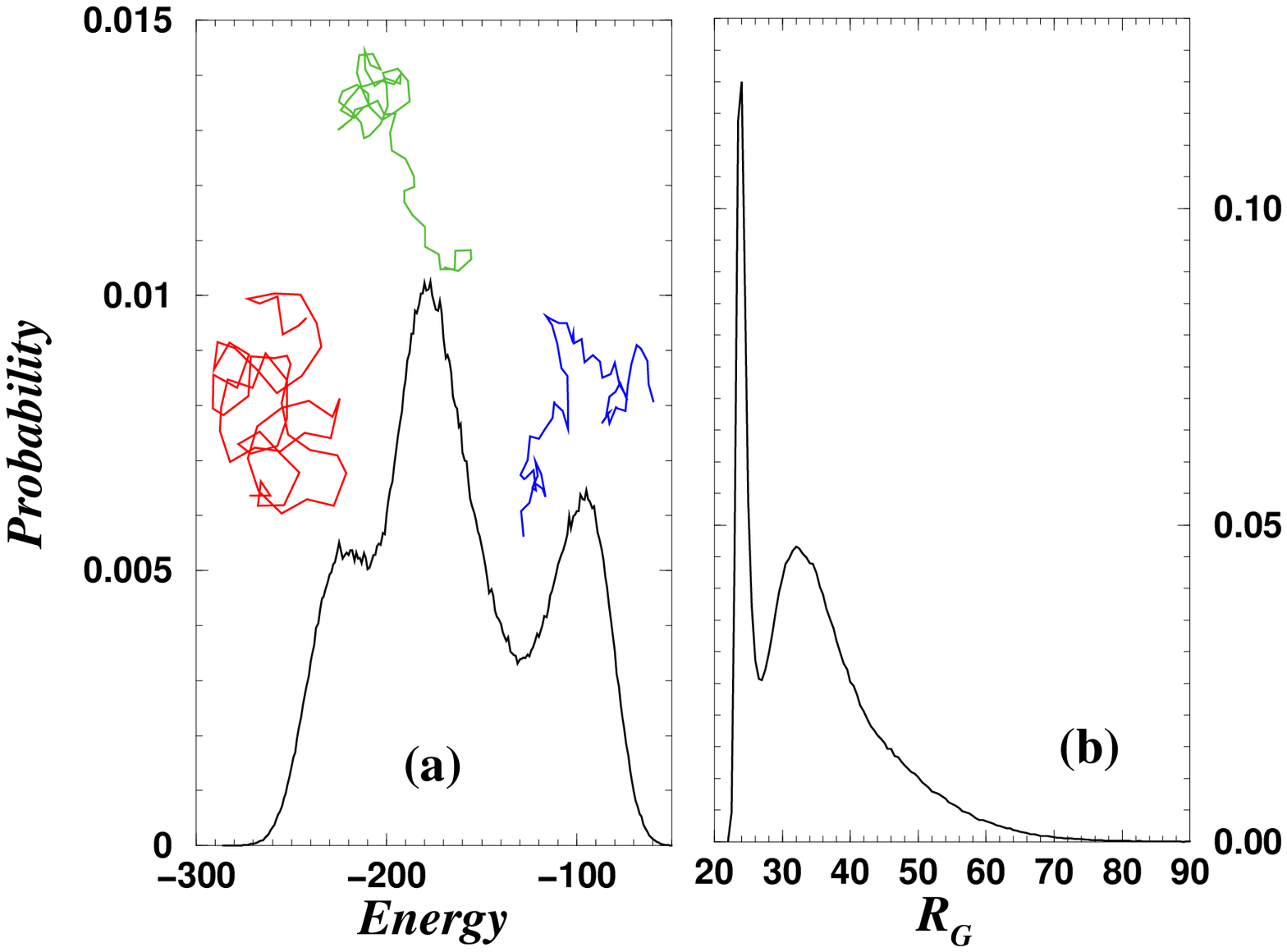} 
\vspace*{1.0cm}
}
\caption{The probability distribution of (a) the energy states ${\cal E}$
and (b) the radius of gyration $R_G$ of the globule maintained at
$T_f=1.46$ for $10^6$ tu. The trimodal distributions indicate
the presence of three states: the folded state, PFS, and the unfolded
state.}
\label{fig:4}
\end{figure}

\begin{figure}[htb]
\centerline{
\epsfxsize=8.0cm
\epsfbox{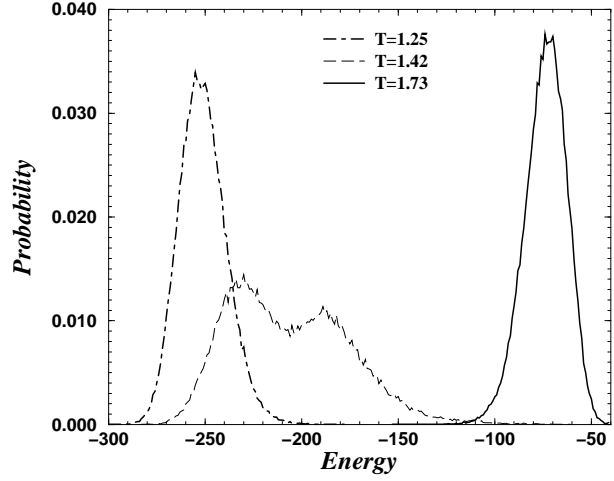} 
\vspace*{1.0cm}
}
\caption{The probability distribution of the energy states ${\cal E}$ 
of the globule maintained at three different temperatures: $T=1.25$,
1.42, and 1.73. Note, that at $T=1.42\approx T_f$ the
distribution has two expressed peaks. The right peak of this ($T=1.42$)
distribution corresponds to the PFS, while the left peak corresponds to
the energetic well of the native state.}
\label{fig:5}
\end{figure}

The simulations reveal that the protein undergoes a folding
$\rightleftharpoons$ unfolding transition as we increase the temperature
to the proximity of the \th-temperature $T_{\theta}$, which in this
model is $T_{\theta}\equiv T_f\approx 1.46$. At $T_{\theta}$ the
distribution of energy has three peaks (Fig.~\ref{fig:4}a). The left
peak corresponds  to the folded state, the right peak corresponds to the
unfolded state, and the middle one corresponds to the partially folded
state (PFS), with 19-residue unfolded tail. This trimodality of the
energy distribution is also seen in Fig.~\ref{fig:3}b. The energy
profile at temperature $T=1.42$ (close to $T_{\theta}$) also reflects
these three states. Since $T < T_{\theta}$, only two states are mostly
present on Fig.~\ref{fig:3}b. Thus, the energy distribution has only two
peaks (Fig.~\ref{fig:5}), corresponding to the folded state and the
PFS. Above $T_{\theta}$, the globule starts to explore energetic wells
other than the native well (see Fig.~13 in \cite{Mirny96}).

\begin{figure}[htb]
\centerline{
\epsfxsize=8.0cm
\epsfbox{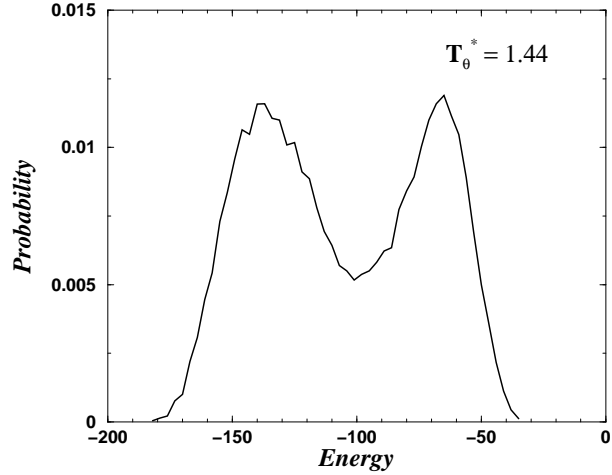} 
\vspace*{1.0cm}
}
\caption{The probability distribution of the energy states ${\cal E}$
of the 46-residue globule maintained at $T_f^{*}=1.44$ during
$10^6$ tu. The bimodal distributions of energy indicates that the
19-residue tail is responsible for the PFS of the 65-residue globule:
after eliminating the 19 residue tail the trimodal energy distribution
of the 65-residue globule becomes bimodal energy distribution of the
46-residue globule.}
\label{fig:6}
\end{figure}

To show that PFS is the cause of the middle peak in energy distribution
(Fig.~\ref{fig:4}a), we eliminate the 19-residue tail and plot for the
46-mer the energy distribution its \th-temperature $T_{\theta}^{*}=1.44$
(Fig.~\ref{fig:6}). We expect to see only two states:
folded and unfolded, since the 19-residue tail, which is the cause of
PFS, is eliminated. Fig.~\ref{fig:6} confirms our expectations.

\begin{figure}[htb]
\centerline{
\epsfxsize=8.0cm
\epsfbox{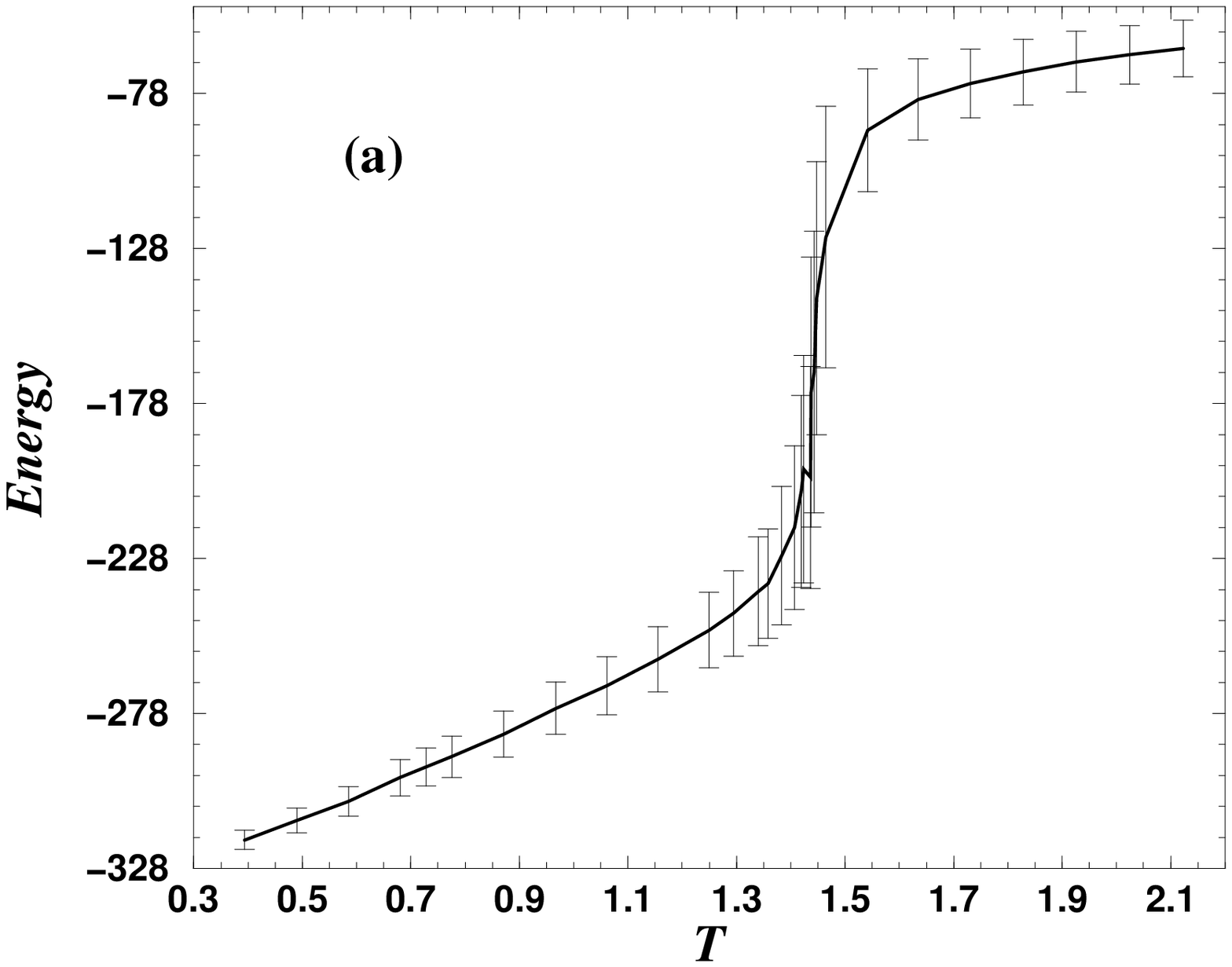} 
\vspace*{1.0cm}
%}
%\centerline{
\epsfxsize=8.0cm
\epsfbox{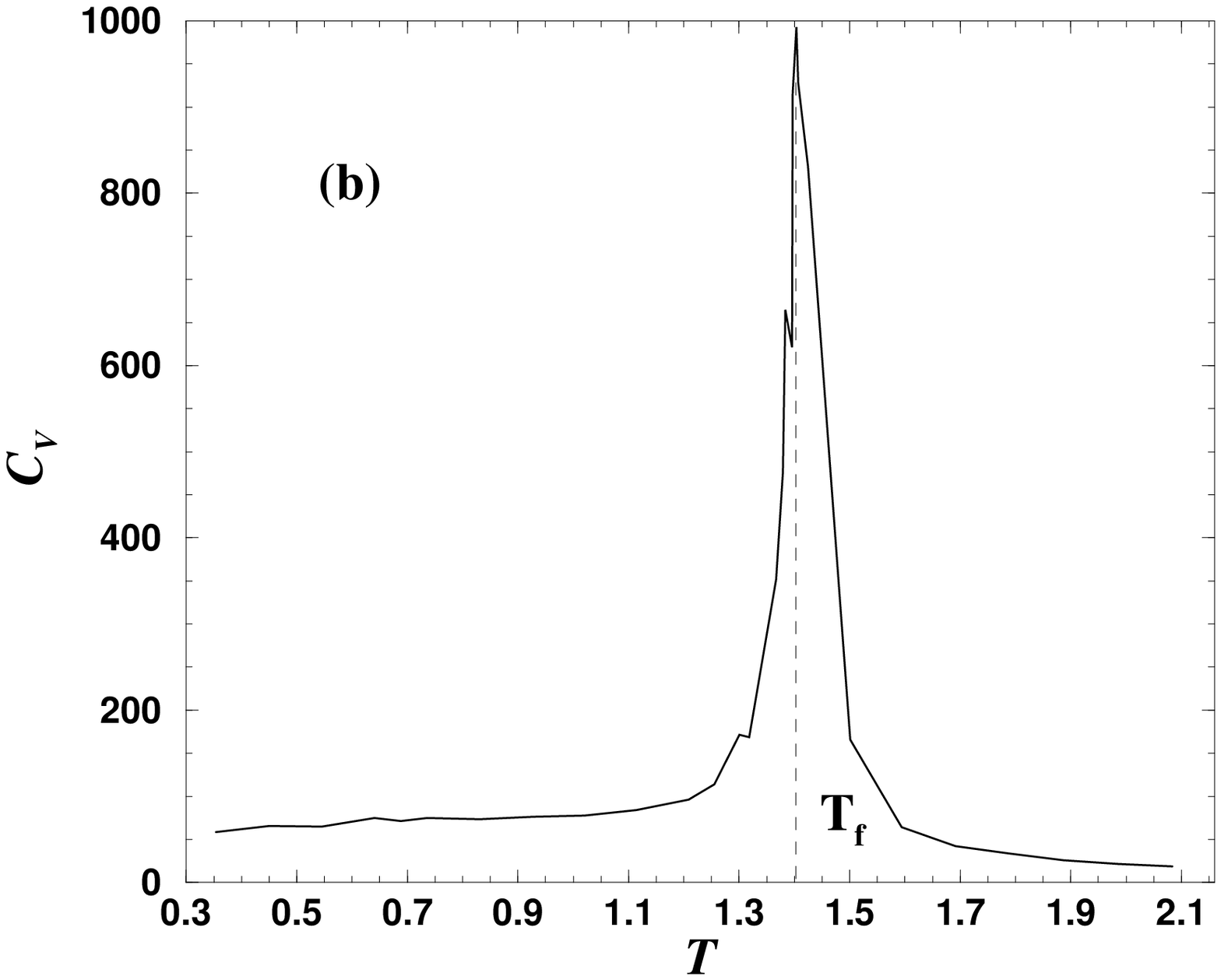} 
\vspace*{1.0cm}
}
\centerline{
\epsfxsize=8.0cm
\epsfbox{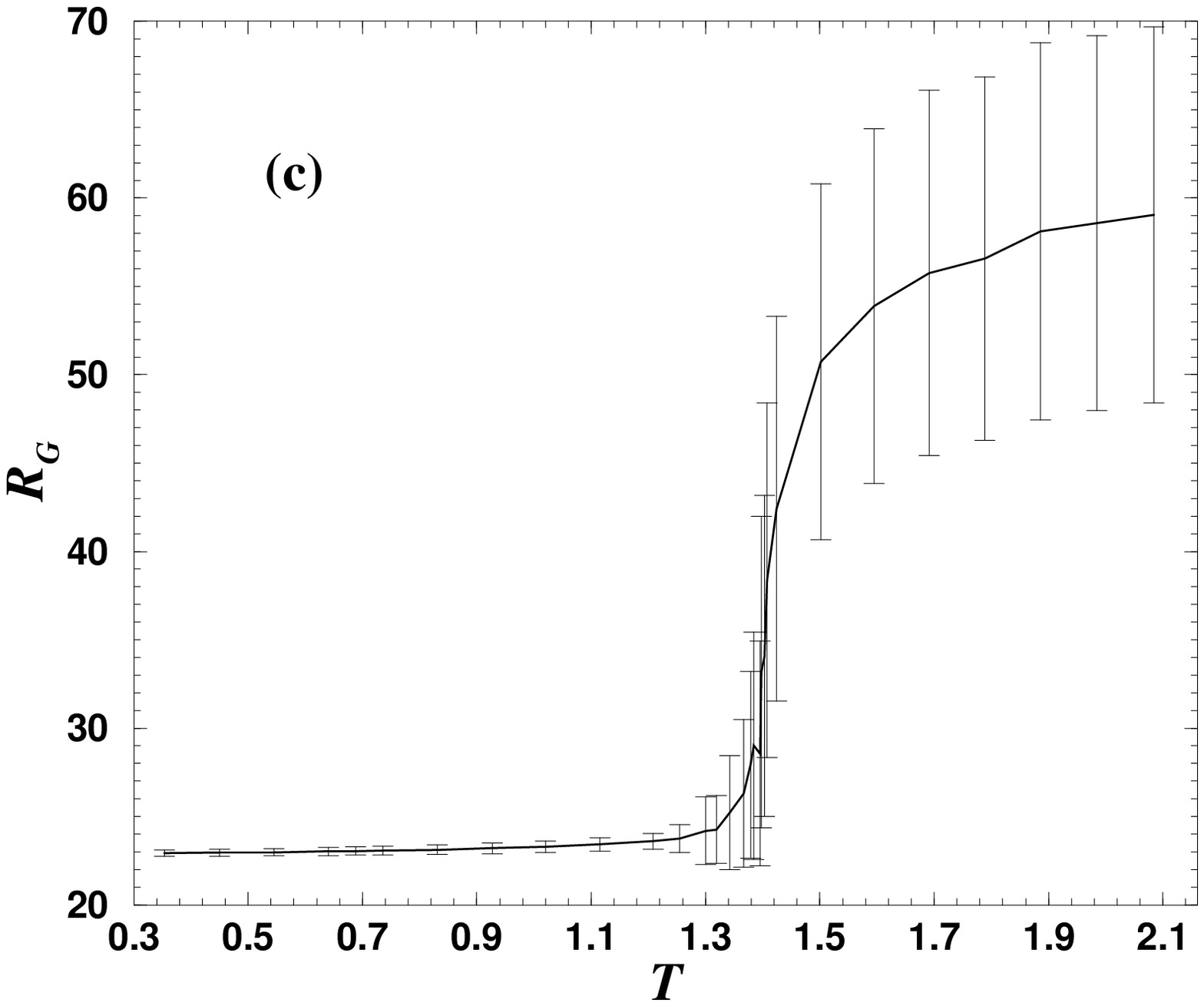} 
\vspace*{1.0cm}
}
\caption{The dependence on temperature of (a) the energy ${\cal E}$, (b)
the heat capacity $C_V$, and (c) the radius of gyration $R_G$. The
error bars are the standard deviation of fluctuations. The rapid
increase of energy as well as the sharp peak in heat capacity at
$T=T_f$ indicates a first order phase transition.}
\label{fig:7}
\end{figure}

\begin{figure}[htb]
\centerline{
\epsfxsize=8.0cm
\epsfbox{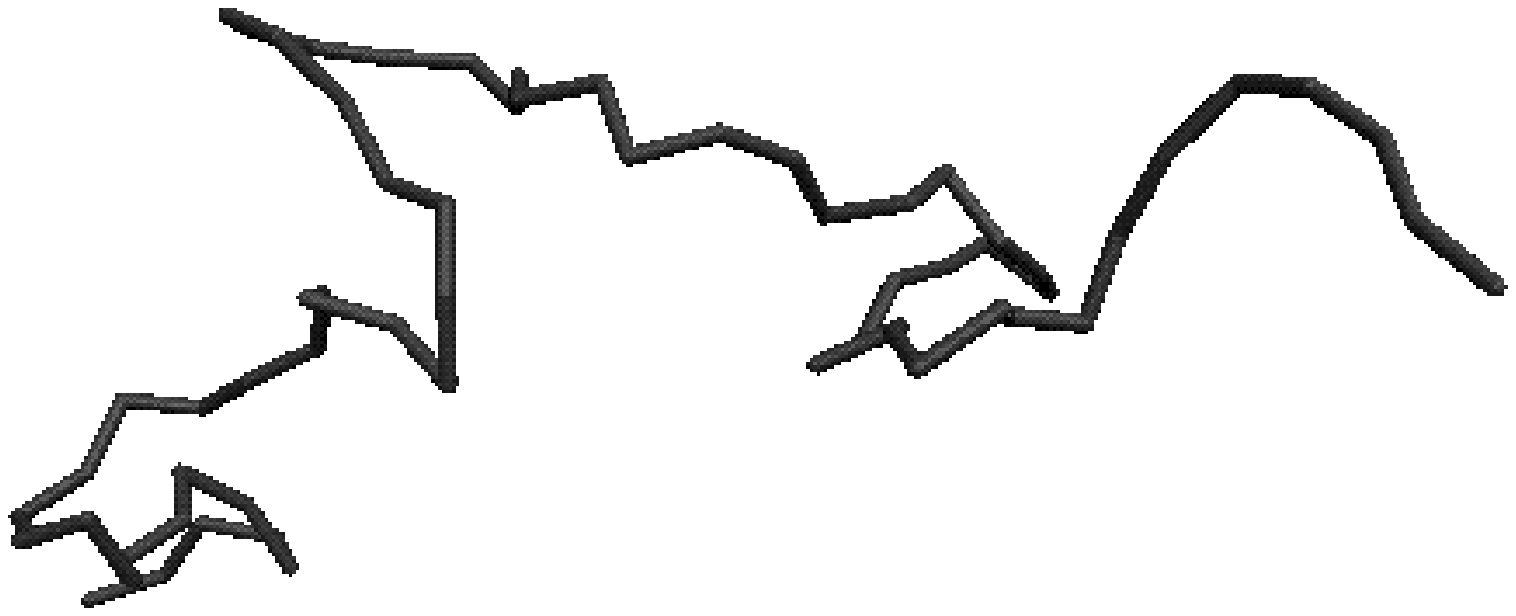}
\vspace*{1.0cm}
%}
%\centerline{
\epsfxsize=8.0cm
\epsfbox{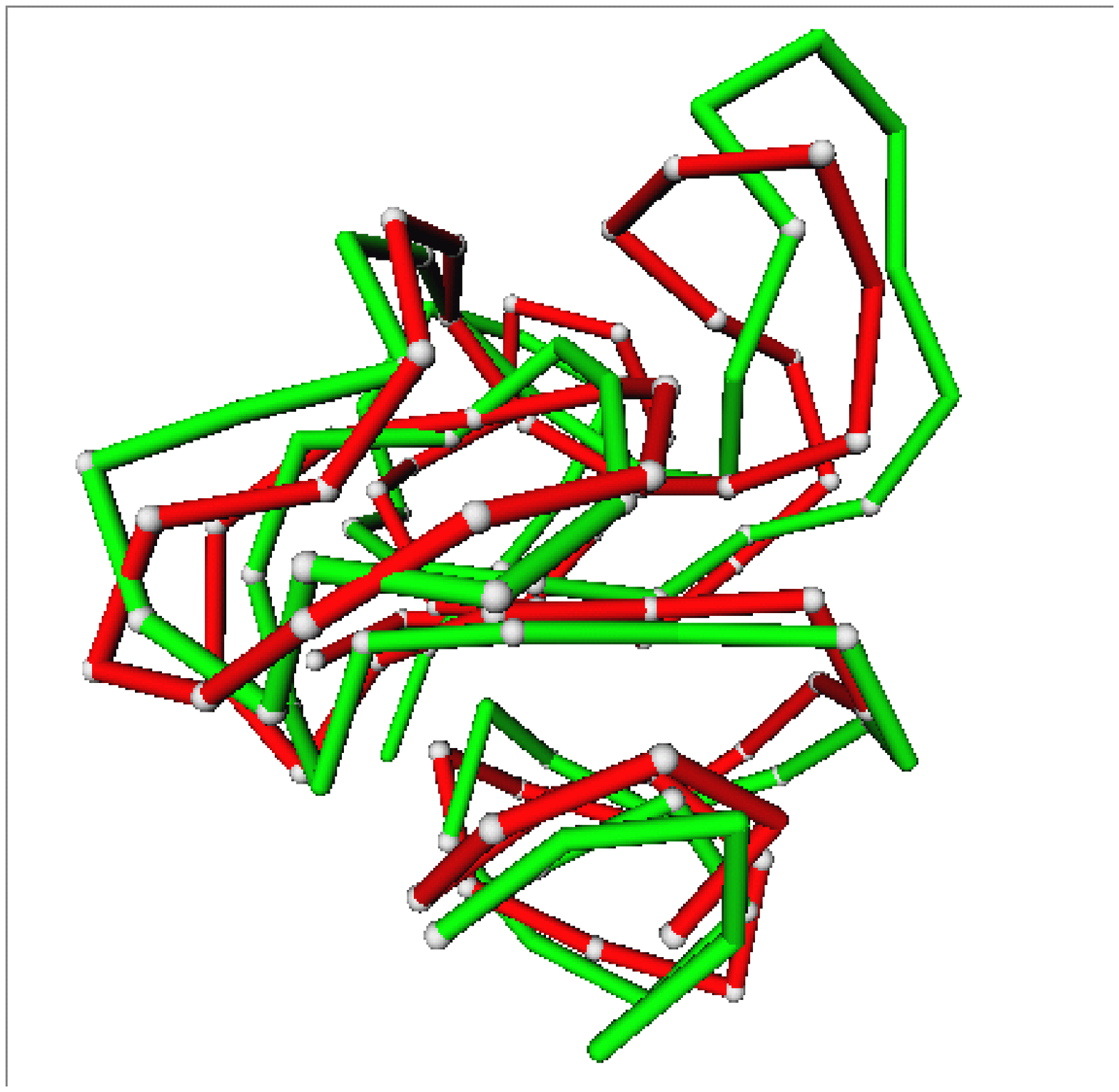} 
\vspace*{1.0cm}
}
\caption{The snapshot of the protein in (a) the unfolded stated, obtained
at high temperature $T=1.8$; and (b) the transition state, obtained
at folding transition temperature $T_f=1.46$ (green), overlapped
with the globule at low temperature $T=0.4$ (red). Note that the TS
globule has a close visual similarity to those maintained at low
temperature and in the native state (see also Fig.~\protect\ref{fig:2}b). It is
more dispersed, however, which makes all the NC easily
breakable. To compare the globule at the TS with the one maintained at
temperature $T=0.4$, we perform the transformation proposed by Kabsch
\protect\cite{Kabsch78} to minimize the relative distance between the
residues in the TS and the state at $T=0.4$. The ``cold'' residues (grey
spheres) denote residues whose rms displacement are smaller than $a_1$.}
\label{fig:8}
\end{figure}

The folding $\rightleftharpoons$ unfolding transition is further
quantified in Fig.~\ref{fig:7}. The energy and the radius of gyration
increase most rapidly near $T_f=T_{\theta}$ resembling the order
parameter jump in a phase transition (see discussion below). This rapid
increase of ${\cal E}$ and $R_g$ reaches its maximum at \th-point, where
the potential of interaction is compensated by the thermal motion of the
particles. Above $T_{\theta}$, interactions between residues do not hold
them together any more and the chain becomes unfolded (see
Fig.~\ref{fig:8}a). Note, that since all the attractive interactions are
specific, the transition is described by one temperature $T_f$.

The presence of the PFS is observed in the temperature range between
1.40 and 1.48, in which the collapse transition occurs. Thus, in this
particular region the folding temperature and the $\theta$-temperature
are indistinguishable within the accuracy of their definitions.

Remarkably, a simple Flory type model of an excluded volume chain predicts
$T_{\theta}$ within 20\%. To demonstrate this, let us write the
probability that the end-to-end distance of the chain is $R$ \cite{Doi96}:
\be
P(R)\propto p(R) \exp(-\frac{N^2 v}{2R^3}-\frac{{\cal  E}(R)}{T})\, 
, \label{eq:P_R}
\ee
where $v=(4\pi/3) (a_0/2)^3$ is the volume of the monomer and
$p(R)\propto R^2\exp(-3R^2/(2N(a_0/2)^2))$ is the probability that the
end-to-end distance of the chain is $R$ for the random walk model. For
$T=T_f$, the repulsive excluded volume term $-(N^2 v)/(2R^3)$
balances the attractive term $-{\cal E}(R)/T>0$. Thus,
\be
T_f = \frac{2 R^3|{\cal  E}|}{N^2 v} \approx 1.7\, ,
\ee
where $E\simeq -130$ and $R\simeq 24$ are taken for a certain
configuration at the \th-point.

We also compute the heat capacity $C_V$ from the relation
\cite{Landau80}:
\be
C_V = \frac{\langle (\delta {\cal E})^2\rangle}{T^2}\, ,
\ee
where $\langle (\delta {\cal E})^2\rangle \equiv \langle {\cal E}^2\rangle -
\langle {\cal E}\rangle^2$ and $\langle ...\rangle$ denotes a time
average. The time average is computed over $10^6$~tu of equilibration at
a fixed temperature. The dependence of the heat capacity on temperature
is shown in Fig.~\ref{fig:7}b. There is a pronounced peak of $C_V(T)$
for $T=T_f$.

We note that below the folding temperature $T_f$ the globule
(Fig.~\ref{fig:8}b) spends time in the state structurally similar to the
native state (Fig.~\ref{fig:2}b). However, one can see that even though
the globule maintains approximately the same structure, i.e. the same
set of NC, the distances between residues are much larger than in the
native state. Due to the fact that the potential of interaction between
like residues is a square-well, there is no penalty for these residues
to be maximally separated, yet they remain within the range of
attractive interaction. This allows the globule to have more NNC and,
thus, still maintain its similar to native structure, yet to have energy
larger than the energy of the native state. This structure can be
identified as the highest in energy, which still maintains its core. As
temperature increases, the ratio $|R_G-R_G^{NS}|/R_G^{NS}$ increases
until temperature reaches $T=T_f$, where the ratio becomes roughly
0.87.

To confirm the presence of the core, we calculate $f\equiv
N_{NC}/N_{C}$, at temperatures below $T=T_f$. The attractive
inter-residue interaction term $-{\cal E}/T$ dominates the excluded
volume repulsion term $-N^2 v/(2R^3)$ (see Eq.~(\ref{eq:P_R})), so
\be
-\frac{{\cal  E}}{T_f} -\frac{N^2 v}{2R^3} >0\, .\label{eq:e_in}
\ee
The total energy $\cal E$ has contributions from both NC and NNC
contacts, so
\be
{\cal  E} = -\epsilon (N_{NC}-N_{NNC})= -[2f-1] \epsilon N_{C} \, .
\label{eq:Energy}
\ee

At a temperature slightly below $T_f$, $|T-T_f|/T_f \approx 0.3$, the
residues are maximally separated within their potential wells, yet they
still maintain contacts. Therefore, the volume $\tilde{v}$ spanned by
one residue is roughly $\tilde{v} \approx (4\pi/3) (a_1/2)^3 =
8v$. $N_C$ is the product of the probability $\tilde{v}/R^3$ of having a
bond (NC or NNC) and the total number of possible arrangements of the
pair contacts between $N$ residues, $N(N-1)/2\approx N^2/2$. Thus,
\be
N_C = \frac{N^2}{2R^3} \tilde{v}\, .\label{eq:N_C}
\ee
From Eqs.~(\ref{eq:e_in}) -- (\ref{eq:N_C}) we can estimate $f$, the
fraction of $N_{C}$ at the temperature $T\approx 1.42 < T_f$:
\be
f> \frac{1}{2} + \frac{v}{\tilde{v}} \frac{T}{\epsilon}\approx
0.68\, .\label{eq:f}
\ee

Due to the fact that the globule maintains roughly the same volume
at temperatures slightly below \th-point, Eq.~(\ref{eq:f}) implies
that approximately 70\% of all native contacts stay intact in the folded
phase (see Fig.~\ref{fig:5}). This
result is supported by the simulations: at $T\approx 1.42$ the number of
NNC is roughly $N_{NNC}\approx 28$, and the energy ${\cal E}$ is ${\cal
E}=-206$. Therefore, the number of NC is $N_{NC}\approx 234$, and the
fraction of NC is $f\approx 0.89$, which is even higher than the lower
limit set by Eq.~(\ref{eq:f}). Note that at a temperature higher than
$T_f$, the fraction of native contacts becomes small due to the fact
that in this regime the interactions are dominated by the excluded
volume repulsion. This change in the number of NC from 70\% to close to
zero indicates the presence of the core structure maintained by these
70\% of NC (see Fig.~\ref{fig:8}b and discussion below). Above the
\th-point the globule is completely unfolded (Fig.~\ref{fig:8}a).

The formation of a specific nucleus during the folding transition was
suggested by many theoretical
\cite{Go83,Shakhnovich97,Abkevich94,Shakhnovich96,Wetlaufer73,Karplus79,Abkevich95,Lazaridis97}
and experimental works \cite{Anifsen73,Tsong78,Bai95,Lacroix97}. The
presence of the core at $T_f$ may arrise from a nucleation processe
driving the system from the unfolded state to the native state. We find
indication of a first order transition. We also offer theoretical
reasoning for the presence of a core (Eq.~(\ref{eq:f})), which might
indicate the presence of a nucleus. Next, we identify the core.

We calculate the mean square displacement $\sigma(T)$ of the
globule at a certain temperature from a globule at the native state,
i.e.
\be
\sigma(T) \equiv \langle [\frac{1}{N} \sum_{i=1}^{N}
(\vec{r}_i^{NS} - \stackrel{\leftrightarrow}{\cal
R}(\stackrel{\leftrightarrow}{\cal T}\vec{r}_i))^2]^{1/2} \rangle
= \langle [\frac{1}{N}\sum_{i=1}^{N} \sigma_i^2(T)]^{1/2} \rangle \, ,
\label{eq:sigma}
\ee
where $\vec{r}_i$ and $\vec{r}_i^{NS}$ are the coordinates of the
residues of the globules at two conformations: at some conformation at
the temperature $T$ and native conformation respectively.
$\stackrel{\leftrightarrow}{\cal T}$ is a translation matrix, which sets
the centers of mass of these configurations at the same point in space.
$\stackrel{\leftrightarrow}{\cal R}$ is a rotation matrix, which
minimizes the relative distance between the residues of two
configurations (for details see
\cite{Kabsch78,Brooks92,Daggett92,Sheinerman97}). The $\sigma_i(T)$ in
Eq.~(\ref{eq:sigma}) are the rms displacements for each individual
residue. 

\begin{figure}[htb]
\centerline{
\epsfxsize=8.0cm
\epsfbox{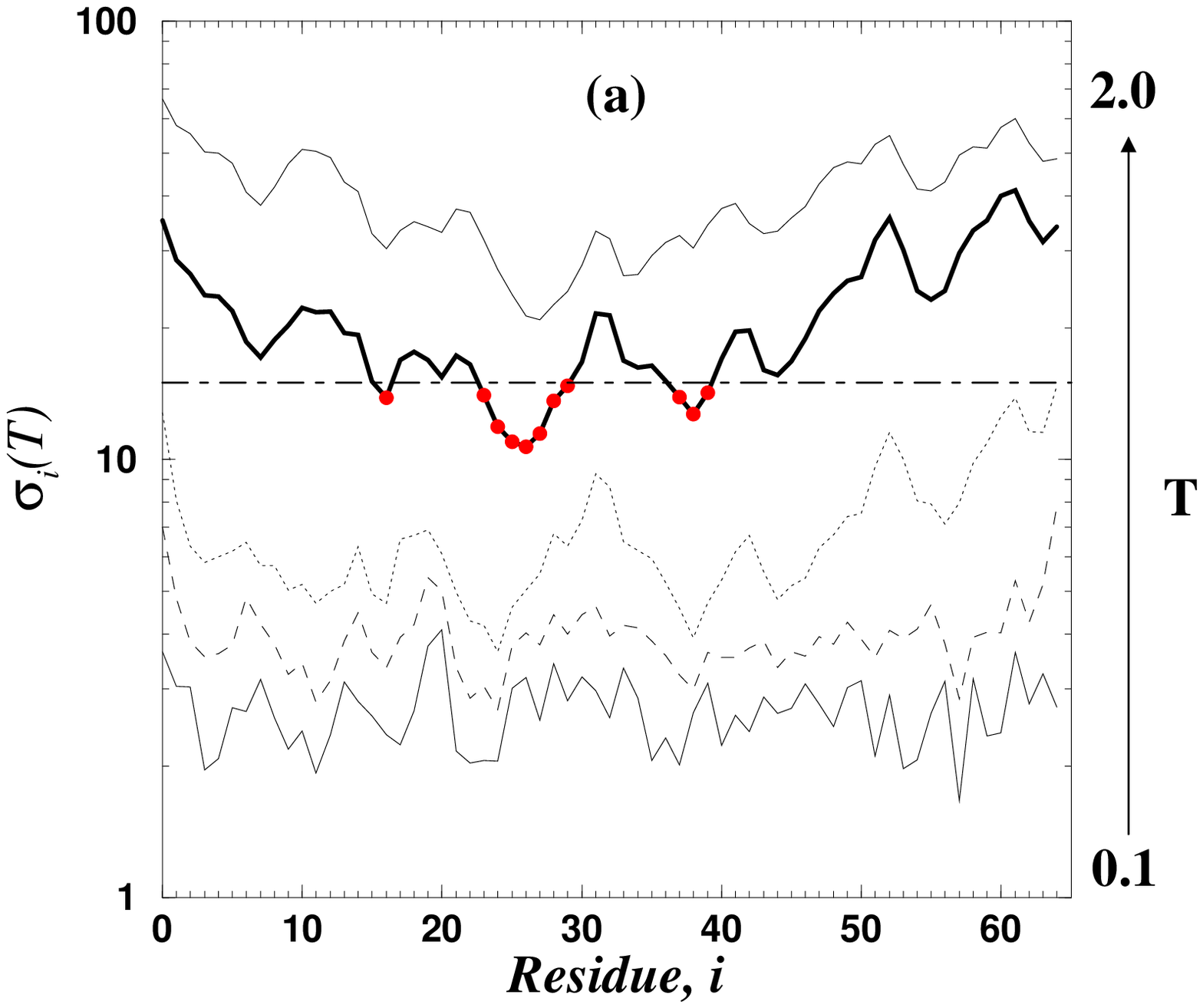} 
\vspace*{1.0cm}
%}
%\centerline{
\epsfxsize=8.0cm
\epsfbox{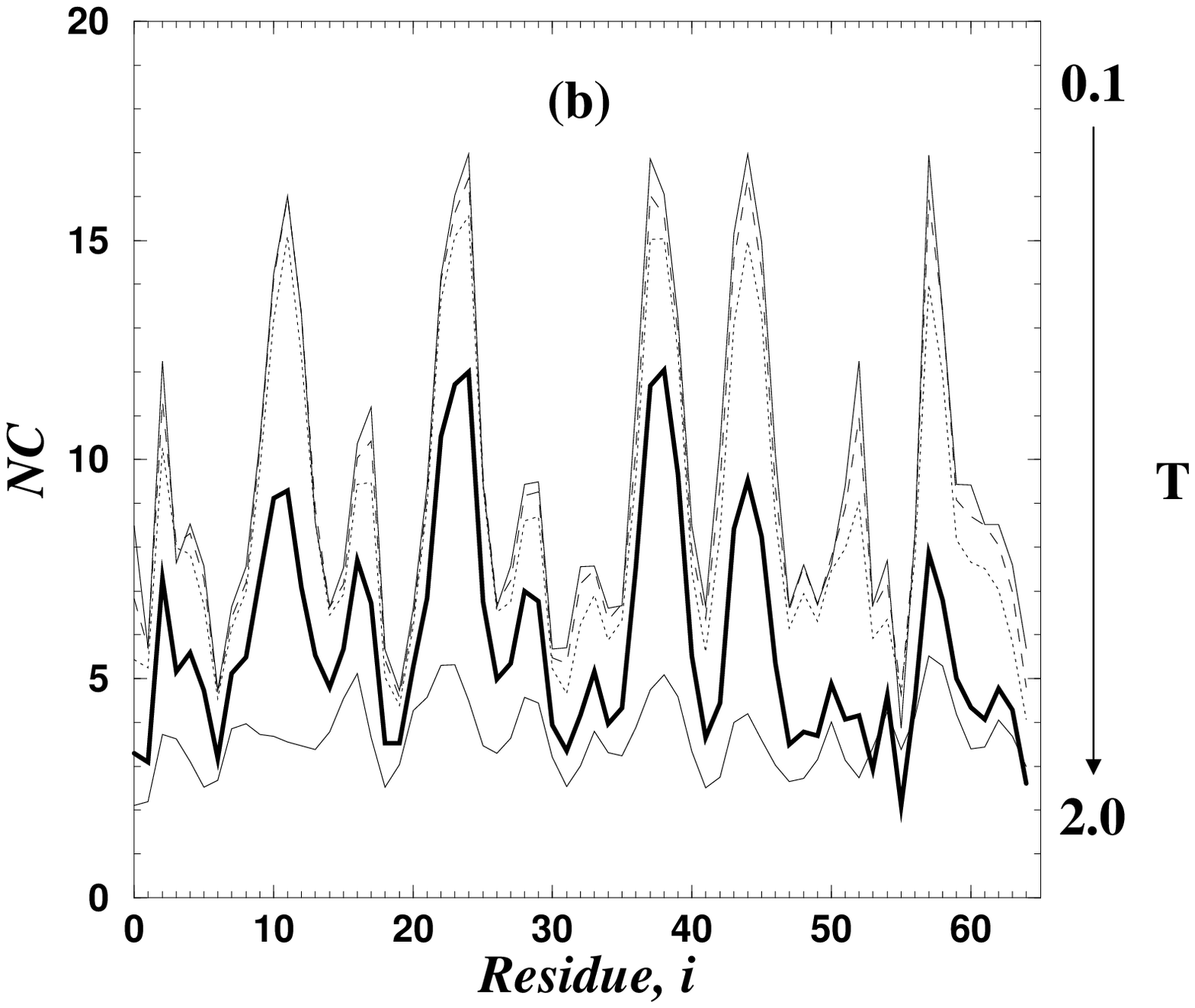} 
\vspace*{1.0cm}
}
\caption{(a) The contour plot of the rms displacement $\sigma_i(T)$ for
each residue $i=0,1,...,64$ at temperatures $T= 0.3$, $0.97$, $1.34$,
$1.46$ (bold line) and $1.54$, averaged over $10^6$ tu. Note that there
is a distinct difference 
between the ``cold'' (small values of $\sigma_i(T)$) and ``hot''
residues (large values of $\sigma_i(T)$). The dashed dotted line
indicates the breaking point of the NC, i.e. when $\sigma_i(T)$
is of the size of the average relative position between pairs of
residues, i.e. $\sigma_i(T) = (a_0+a_1)/2 \approx 15$. The bold lines
(on both (a) and (b)) indicate the folding transition temperature line
$T_f$. It is worth noting that 11 residues are still in contact
(marked by red circles on (a)): 16, 23, 24, 25, 26, 27, 28, 29, 37,
38, 39. (b) The analogous to (a) plot of the average number of NC for each
residue. Note that the number of NC is strongly correlated with the
rms: the local minima of the $\langle\sigma_i(T)\rangle$ plots
correspond to the local maxima of the number of NC.}
\label{fig:9}
\end{figure}

The plot of $\langle\sigma_i(T)\rangle$ is presented in
Fig.~\ref{fig:9}a. From the roughness of the ``landscape'' in
Fig.~\ref{fig:9}a, we can select a group of residues whose rms
displacements are significantly smaller than the rms displacements of
the other group of residues. We denote the former group by ``cold''
residues and the latter group by ``hot'' residues. The rms displacement
strongly depends on the temperature near the folding transition and
grows slowly below $T_f$. Note that the average
numbers of NC of the residues are correlated with the average rms
displacement of these residues, i.e. the peaks on the $N_{NC, i}$
isothermal lines of Fig.~\ref{fig:9}b correspond to the ``cold''
residues.

\begin{figure}[htb]
\centerline{
\epsfxsize=8.0cm
\epsfbox{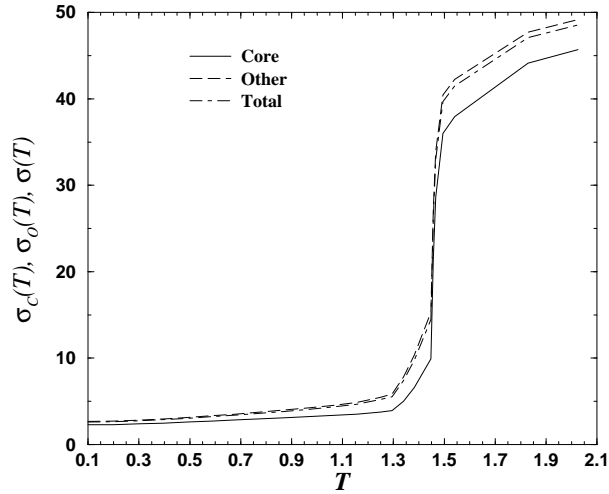} 
\vspace*{1.0cm}
}
\caption{The dependence of the rms displacement of the core residues
$\sigma_C(T)$ (solid line), the rest of the residues $\sigma_O(T)$
(dashed line) and all the residues $\sigma(T)$ on temperature. The above
quantities are averaged over $10^6$ tu. Note, that for the ideal first
order phase transition one would expect $\sigma_C(T)$ to be a
step function. However, since we consider a transition that would be
first order in the limit of the infinite size, $\sigma_C(T)$ exhibits
only step-function like behavior. The difference
between core residues and other residues is that at $T_f$ the average
rms displacement of the core residues is smaller than 15, which
indicates that they are in contact (see caption to the
Fig.~\protect\ref{fig:9}). On the contrary, the average rms displacement
of the non-core residues is greater than 15, indicating that these
residues are not in contact.}
\label{fig:10}
\end{figure}

Next, we calculate the rms displacement $\sigma_C(T)$ for the selected
25\% coldest residues (the core) and $\sigma_O(T)$ for the rest of the
residues. Fig.~\ref{fig:10} shows their dependence on temperature, as
well as the dependence of the rms displacement for all residues
$\sigma(T)$. There is a pronounced difference in the behavior of
the rms displacement of the core residues and the rest of the residues
below $T_f$. At $T_f$ their behavior is the same, due to the fact that
all the attractive interactions are balanced by the repulsion of the
excluded volume. Above $T_f$ the difference between $\sigma_C(T)$ and
$\sigma_O(T)$ is only due to the fact that the core residues have most
of the NC and, therefore, are more likely to spend time together even at
$T > T_f$.

\begin{figure}[htb]
\centerline{
\epsfxsize=8.0cm
\epsfbox{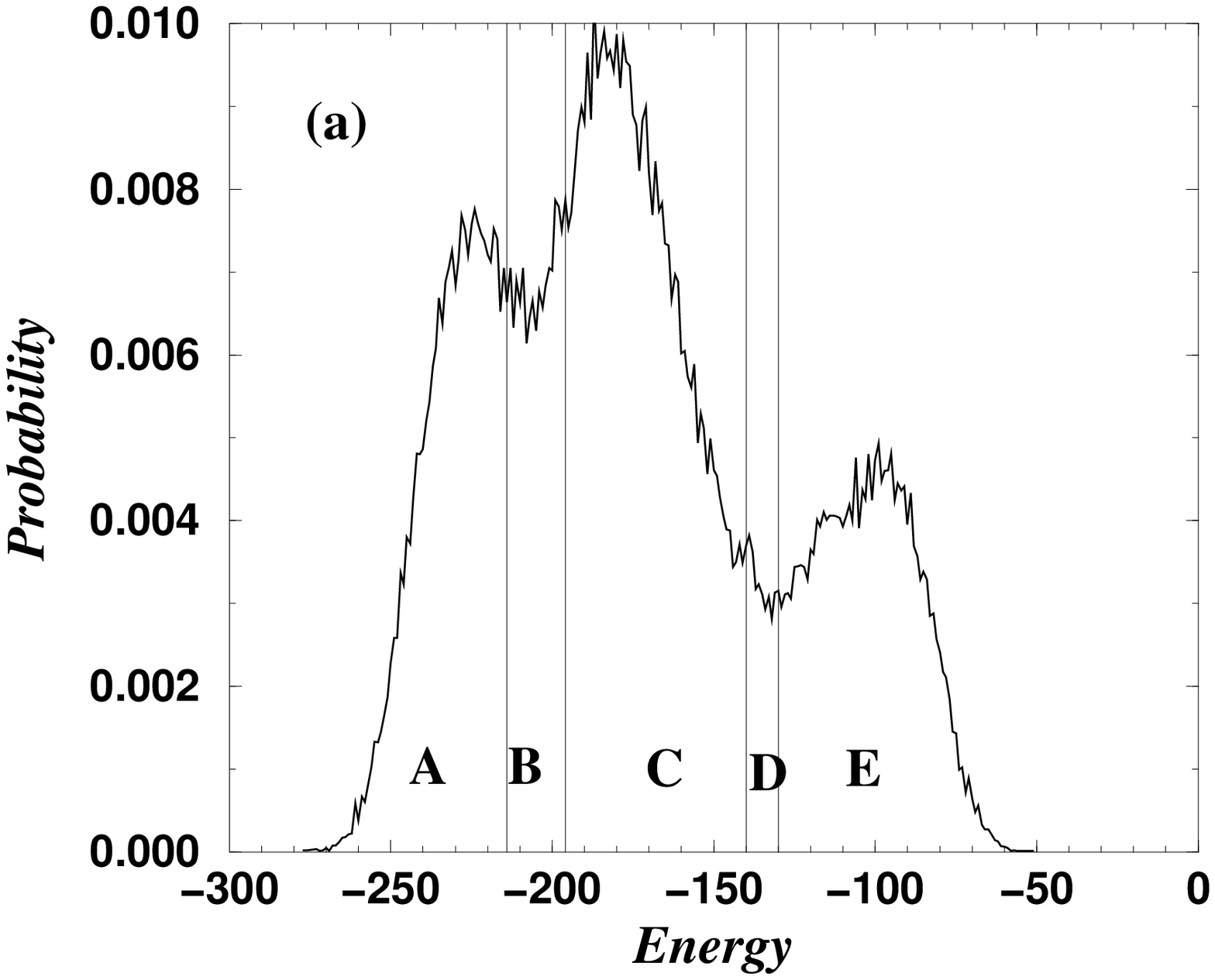} 
\vspace*{1.0cm}
%}
%\centerline{
\epsfxsize=8.0cm
\epsfbox{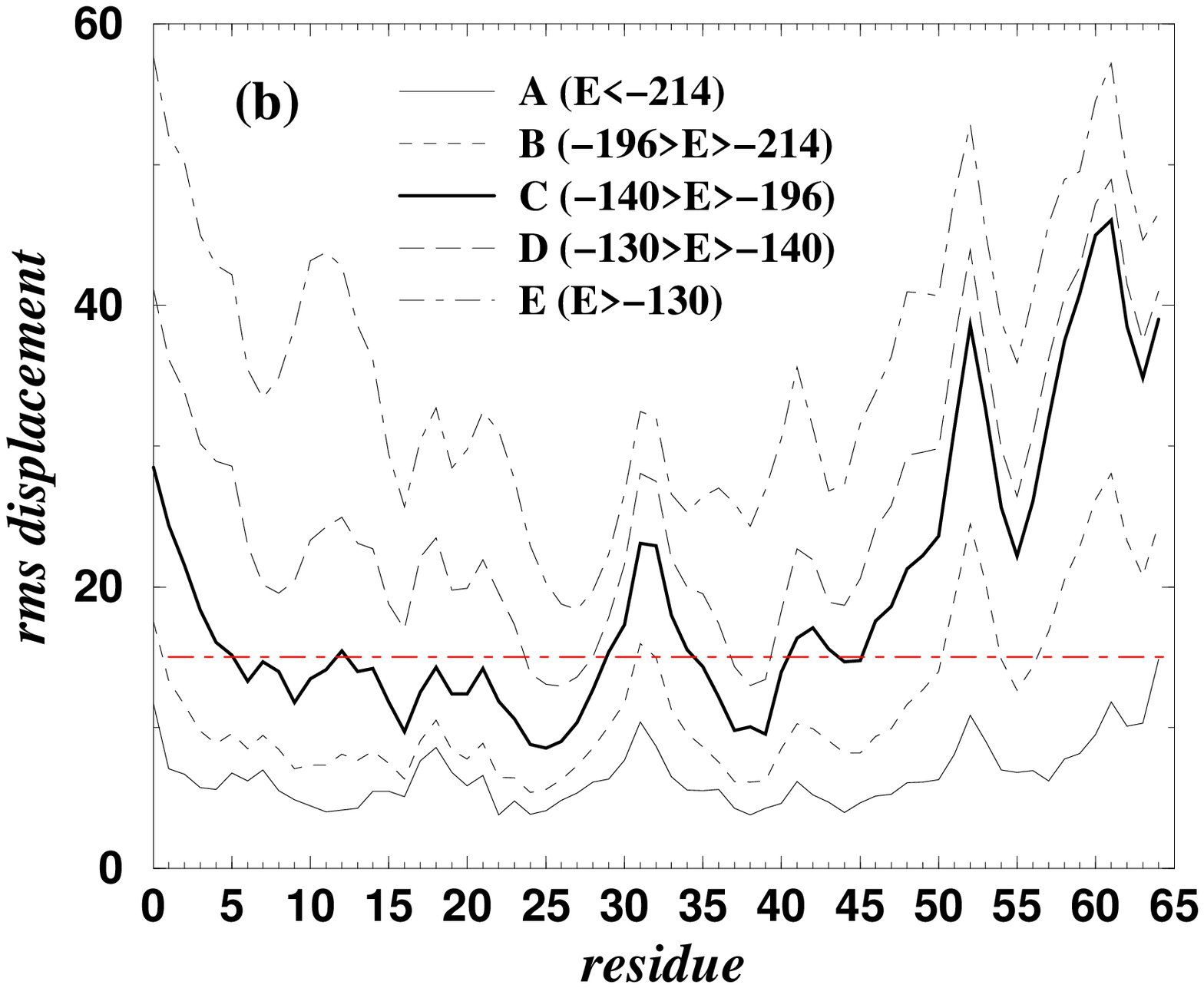} 
\vspace*{1.0cm}
}
\centerline{
\epsfxsize=8.0cm
\epsfbox{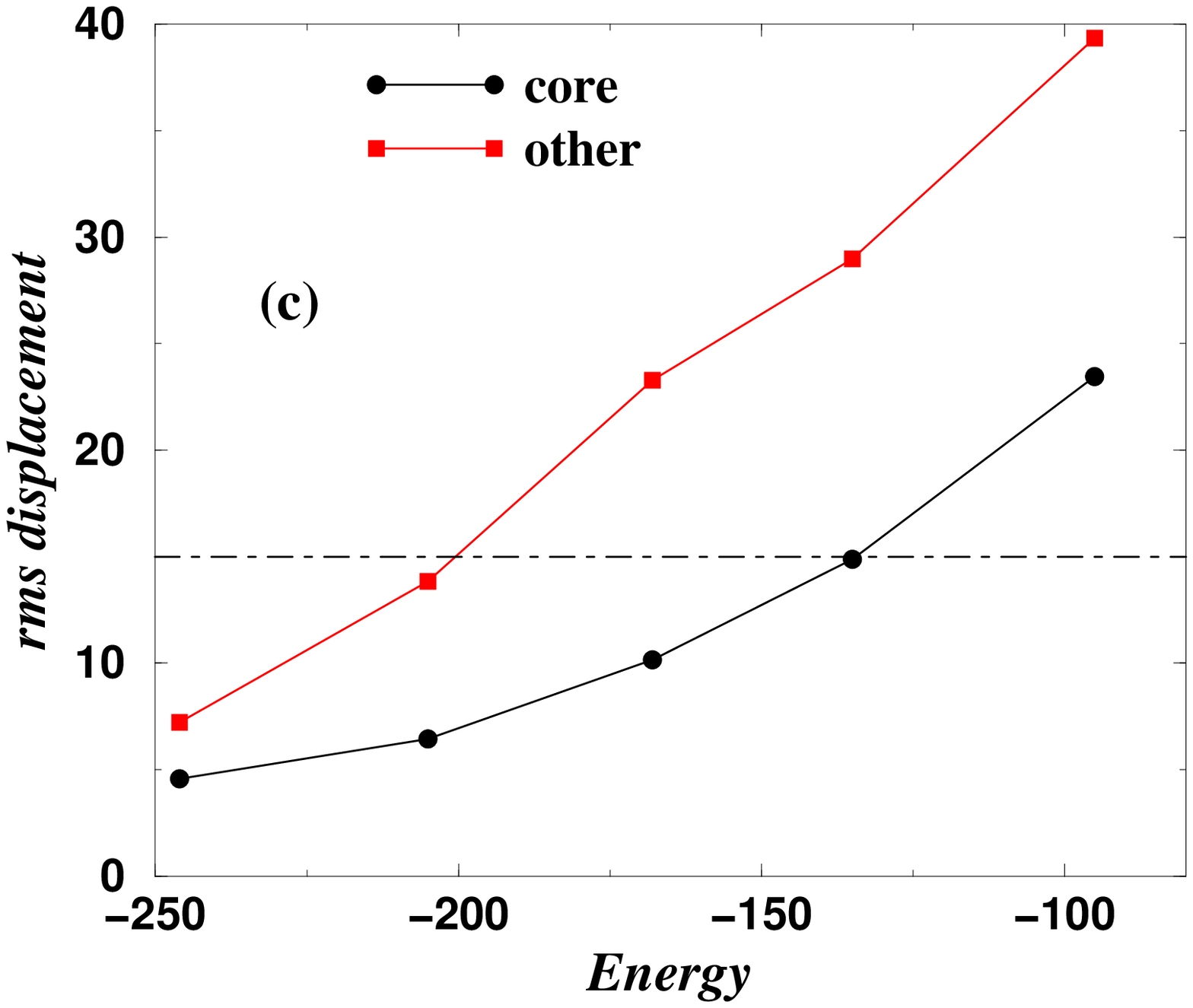} 
\vspace*{1.0cm}
}
\caption{(a) The probability distribution of the energy states ${\cal E}$
of the globule maintained at $T_f=1.46$ during $10^6$ tu. The
probability distribution is divided into five regions: $A$, $B$, $C$,
$D$, and $E$. Region $A$ corresponds to the folded state; region $B$
corresponds to the transitional state between folded state and PFS;
region $C$ corresponds to the PFS; region $D$ corresponds to the
transitional state between PFS and completely unfolded state. (b) The
plot of the rms displacement $\sigma_i(T)$ for each residue
$i=0,1,...,64$ for various regions $A$, $B$, $C$, $D$, and $E$ of the
plot (a) averaged over $10^6$ tu. Note, that in region $A$ all residues
stay in contact; in region $C$ both N- and C-terminus tails break away,
forming PFS; in region $D$ there are only a few of core residues are
still stay intact; and in region $E$ none of the residues are in
contact. (c) The dependence of the rms displacement of the core residues
(circles), the rest of the residues (squares) on the average energy
${\cal E}$ of the window of the corresponding region. Note, that core
residues stay intact even in the second transitional state $D$ between
the PFS and completely unfolded state. The dashed dotted line in  (b)
and (c) indicates the breaking point of the NC (see
Fig~\protect\ref{fig:9}).}
\label{fig:11}
\end{figure}

To study the behavior of the globule at $T_f$, we subdivide the 
probability distribution of the energy states ${\cal E}$ of the globule
maintained at $T_f=1.46$ during $10^6$ tu into five regions: $A$,
$B$, $C$, $D$, and $E$ (see Fig.~\ref{fig:11}a). Region $A$ corresponds
to the folded state; region $B$ corresponds to the transitional state
between folded state and PFS; region $C$ corresponds to the PFS; region
$D$ corresponds to the transitional state between PFS and completely
unfolded state. Next we plot the rms displacement for each residue
for each of the above regions (see Fig.~\ref{fig:11}a). Note, that in
region $A$ all residues stay in contact; in region $C$ both N- and
C-termini tails break away, forming PFS; in region $D$, there are only a
few core residues that still stay intact; and in region $E$ none of
the residues is in contact. In region $B$, we observe that part of the
C-terminus tail residues are not in contact, indicating the formation of
a PFS. Next, we plot the dependence of the selected 11 core residues
(see caption to Fig.~\ref{fig:9}) on the average energy of the window of
the corresponding region (see Fig.~\ref{fig:11}c). We observe that core
residues remain close to one another even in the second transitional
state $D$ between the PFS and completely unfolded state.

\begin{figure}[htb]
\centerline{
\epsfxsize=8.0cm
\epsfbox{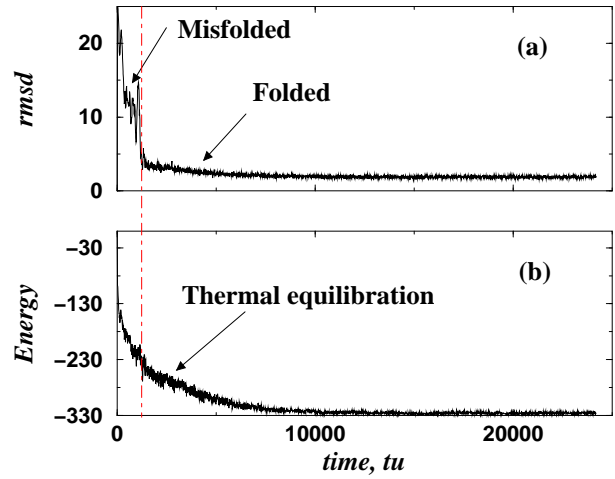} 
\vspace*{1.0cm}
}
\caption{The time evolution of the (a) rms displacement per residue of the
globule from its native state and (b) energy when we cool the system
from the high temperature ($T=3.0$), unfolded state down to the low
temperature ($T=0.1$) state. The model protein gets trapped in the
misfolded conformation after 200~tu and then proceeds to its native
state after 1000~tu. Although the rms displacement of the globule from
its native state is close to 0 after 1200~tu the energy of the globule
is higher than energy of the native state for about $10^4$~tu. This
effect is due to the thermal bath ghost particles which thermally
equilibrate the system during $t_{relax.}\protect\approx 10^4$~tu
relaxation time.}
\label{fig:12}
\end{figure}

We also study the system by cooling it from the high temperature
state. This technique corresponds to the simulated annealing, due to the
fact that the temperature control is governed by the ghost particles
that are present in the system. We find that if the target temperature
is above 1.1 the globule always reaches the state corresponding to
native state. However, if the target temperature is 0.96, the globule
reaches the state, corresponding the native state only in $\approx 70\%$
of the cases, in the time interval of $10^5$ time units. As an example
we demonstrate on Fig.~\ref{fig:12} the cooling of the model protein
from the high temperature state $T=3.0$ to the low temperature state
$T=0.1$. The model protein collapses after 1200~tu.

What is particularly remarkable about Fig.~\ref{fig:12} is that we can
follow the kinetics of the collapse. First, the globule gets trapped in
some misfolded conformation, where it stays for about 1000~tu (see
Fig.~\ref{fig:12}a), and then it collapses to the native state. The time
behavior of the energy, however, can look a bit puzzling. After the rms
displacement drops to close to 0, indicating the native state, the
energy is still higher than that of the native state for about $10^4$~tu
(see Fig.~\ref{fig:12}b). The key to resolve this puzzle is the fact
that after the collapse of the model protein its potential energy
transforms to kinetic energy, which slowly decreases by thermal
equilibration with the bath of the ghost particles.

\section{Discussion}

We find that the classical model of the self-avoiding chain with
excluded volume shows good agreement with the simulations. We show
from simple arguments and simulations that the fraction of NC at the
folding temperature $T_f$ is larger than 70\%, consistent with the
presence of the core. The nucleus forms in the unstable transition
state. From the transition state the globule jumps either to the
completely unfolded conformation or to the folded conformation.

Our simulations are in agreement with the recent work of
Zhou and Karplus \cite{Zhou97a}. They performed discrete molecular
dynamics simulations of {\em Staphylococcus aureus} protein A, the
inter-residue interactions of which were modeled based on G\bo model
\cite{Taketomi75,Go81,Abe81}. The pair residues of model protein, which
form native contacts, had ``square-well'' potential of interaction with
the depth of the well equal to $B_N\epsilon$, while all other pair
residues had ``square-well'' potential of interaction with the depth of
the well equal to $B_O\epsilon$. They characterized the difference
between NC and NNC by the {\em ``bias gap''}, $g$: $g=1-B_O/B_N$. Zhou
and Karplus found that when $g=1.3$, i.~e. when the interaction between
NC is of the opposite sign to the interaction between NNC, there is a
strong first-order-like transition from the random coil to the ordered
globule. The case with our globule corresponds to $g=2$, where,
according to the work of Zhou and Karplus there should exist a strong
first-order-like transition from the random coil to the ordered globule
without intermediate.

We also select the core residues and show that their rms displacement
behaves significantly differently than the behavior of the rms
displacement of the rest of the residues and exhibits step-function like
behavior upon the change of temperature. Our findings are in agreement
with the recent experimental study of the equilibrium hydrogen exchange
behavior of {\it cytochrome c} of Bai et al. \cite{Bai95}, who
investigated the exposure of the amide hydrogens (NH) in {\it cytochrome
c} to solvent (due to local and global unfolding fluctuations). The
experiments were based on the properties of the amide hydrogens that are
involved in hydrogen-bonded structure and can exchange with solvent
hydrogens. Bai et al. demonstrated that proteins undergo folding
$\rightleftharpoons$ unfolding transition ``...through intermediate
forms''. They also selected these intermediate forms (cooperative
units), which are 15 to 25 residues in size. The presence of PFS in our
simulations is thus in agreement with the findings of Bai et al. of the
intermediate forms in {\it cytochrome c}.

The relation between core residues that we find and the nucleus is hard
to establish due to the fact that TS is very unstable. Recent amide
hydrogen exchange experiments on CheY protein from {\it Escherichia
coli} of Lacroix et al. \cite{Lacroix97} provided the 
evidence for the residues involved in the folding nucleus. Furthermore,
the lattice MC simulations of Abkevich et al. \cite{Abkevich94} also
demonstrate that the presence of the nucleus is a necessary and
sufficient condition for subsequent rapid folding to the native state. The crucial
difference between the nucleus and the rest of the structure is in
dynamics, which is manifest also in equilibrium fluctuations. All local
unfolding fluctuations (i.e. the ones after which the chain returns
rapidly back to the native state) keep the nucleus intact, while
fluctuations that disrupt the nucleus lead to global unfolding:
``descend'' to the ``unfolded'' free energy minimum
\cite{Shakhnovich97,Abkevich94}. This view is consistent with the
hydrogen exchange experiments \cite{Bai95,Lacroix97}. Such behavior of
the globule is consistent with a possible first order phase transition
in a system of finite size.

\section{Acknowledgments}
We would like thank G. F. Berriz for designing the globule,
Dr. V. I. Abkevich, L. Mirny, M. R. Sadr-Lahijani, Prof. S. Erramilli
for helpful discussions, and  R. S. Dokholyan for help in editing the
manuscript. N. V. D. is supported by NIH NRSA molecular biophysics
predoctoral traineeship (GM08291-09). E. I. S. is supported by NIH grant
RO1-52126.

%\section{Legends}

\end{document}